# Broadband Setup for Magnetic-Field-Induced Domain Wall Motion in Cylindrical Nanowires

Alexis Wartelle[1,2], Christophe Thirion[1,2], Raja Afid[1,2], Ségolène Jamet[1,2], Sandrine Da Col[1,2], Laurent Cagnon[1,2], Jean-Christophe Toussaint[1,2], Julien Bachmann[3], Sebastian Bochmann[3], Andrea Locatelli[4], Tevfik Onur Menteş[4], and Olivier Fruchart[1,2]

[1]Université Grenoble Alpes, Institute Néel, Grenoble F-38000, France
[2]Institute Néel, Centre National de La Recherche Scientifique (CNRS), Grenoble F-38000, France
[3]Friedrich–Alexander Universität Erlangen–Nürnberg, Erlangen 91058, Germany
[4]Elettra–Sincrotrone Trieste S.C.p.A., Trieste I-34012, Italy

**In order to improve the precision of domain wall (DW) dynamics measurements, we develop a coplanar waveguide-based setup, where the DW motion should be triggered by pulses of magnetic field. The latter are produced by the Oersted field of the waveguide as a current pulse travels toward its termination, where it is dissipated. Our objective is to eliminate a source of bias in DW speed estimation while optimizing the field amplitude. Here, we present the implementations of this concept for magnetic force microscopy and synchrotron-based investigation.**

## I. INTRODUCTION

MAGNETIC domain walls (DWs) have been discovered almost a century ago, yet their investigation has triggered a large research effort in recent years. This interest is driven by hopes for applications using DWs, as well as by the progress made in fabrication allowing a great diversity of nanostructures, leading to novel physics. At the nanoscale, the geometry influence on the equilibrium magnetic configuration is very strong, and leads to several types of DWs, as shown in [1] and [2]. Predicting and confirming [3] the DW type is crucial, because the dynamic properties are heavily affected by the equilibrium configuration. For instance, vortex DWs and transverse DWs are expected to quickly enter a precessional, low-speed (∼100 m/s) regime under field [4], while a Bloch point DW [3] has been predicted [1] to reach the speeds up to a few kilometers per second.

The difficulty in unraveling the details of a DW's dynamics lies in the fact that only a small number of techniques, such as stroboscopic X-ray magnetic circular dichroism (XMCD), can produce time-resolved imaging of DW structures during their motion, provided that the initial magnetic state and the subsequent motion can be repeated a very large number of times [5]. Another approach consists in performing static imaging of DWs before and after applying a pulse of magnetic field (or spin-polarized current). The ratio of the traveled distance to the pulse duration yields an estimate of the average DW speed, which can be compared with the simulations. However, if the device used for the field generation allows a partial pulse power reflection at its end, there is a magnetic field echo that may induce an additional DW displacement. Speed measurements are thus biased.

We report here on the development of a method for the investigation of the dynamics of a DW, where cylindrical nanowires are placed on top of a coplanar waveguide (CPW) using focused ion-beam (FIB)-based micromanipulation. The CPW is tuned to optimize the generation of a magnetic field pulse on a nanosecond time scale. The use of a matching impedance at the end of the device dissipates the pulse, so that none of its power can be reflected, thus eliminating one possible bias. Adaptation of this method for XMCD-photoemission electron microscopy (PEEM) and for magnetic force microscopy (MFM) are presented.

## II. MATCHED COPLANAR WAVEGUIDE

Our purpose is to fabricate a waveguide with a 50 Ω impedance matching. Starting from the voltage pulse generator (model AVG-4B-C by Avtech, pulsewidth at a half-maximum of 3.5 ns) whose output is matched to 50 Ω, we use a certain type of coaxial cables of 50 Ω characteristic impedance to deliver pulses to the waveguide with minimal losses and signal reflections. A proper electrical contact to the waveguide is ensured with the probes of our design for the MFM experiments, and with spring-loaded contacts for the synchrotron experiments. Impedance matching along the length of the waveguide is obtained by geometrical conditions, and the termination itself is a $Ni_{80}Cr_{20}$ strip, whose dimensions are adapted to produce a discrete 50 Ω load. This ensures pulse propagation with minimal losses up to the termination, where the pulse power is dissipated. As a result, no pulse echo travels backward, preventing additional fields that would bias the measurement.

### A. Copper–Beryllium Probes for MFM Experiments

MFM is a scanning probe microscopy deriving from atomic force microscopy (AFM), and is sensitive to the stray fields from the sample. It is a slow but a reliable technique, which does not require synchrotron facilities, and is thus more practical. As of now, its sensitivity is not sufficient to resolve the DW structures, but it is possible to monitor the position of DWs (which can be viewed as magnetic charges creating a stray field) along a wire or a strip. We have designed a copper–beryllium probe to convert the SMA coaxial geometry into planar contacts for the CPW [Fig. 1(a)]. It consists of two pieces of annealed copper–beryllium foils fixed to an SMA connector. The smaller piece of the foil is soldered in a notch in the connector's pin, and contacts the central track of the CPW. Ground continuity is ensured by the other piece of foil, which divides into two tongs on the left and on the right of the CPW. In all the three cases, the tips of the tongs are curved, so that contact is ensured along a line. A linear transition from the connector's lateral dimensions to the dimensions of the CPW was chosen for the tongs, with a constraint on the lateral aspect ratio of the tongs for impedance matching, as explained in Section II-B. The probes are well suited for reliable contacting, provided the CPW is long enough to separate this system from the microscope head.

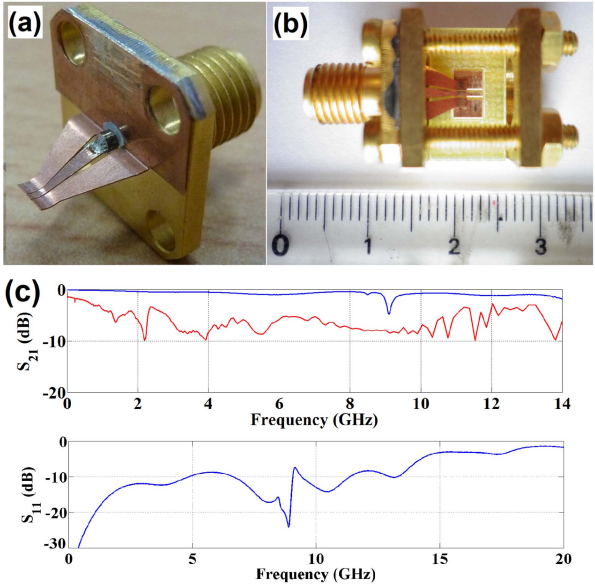

Fig. 1. (a) Copper–beryllium probe on its SMA connector. The tapering from the connector to the tip of the tongs is linear. (b) Setup for HF characterization of the probes; only one is present in order to show the underlying copper strips. (c) $S_{11}$ and $S_{21}$ measurements of the contacting probes (blue lines) describing reflection and transmission properties of the device, as well as $S_{21}$ measurement of a prototype CPW (red line).

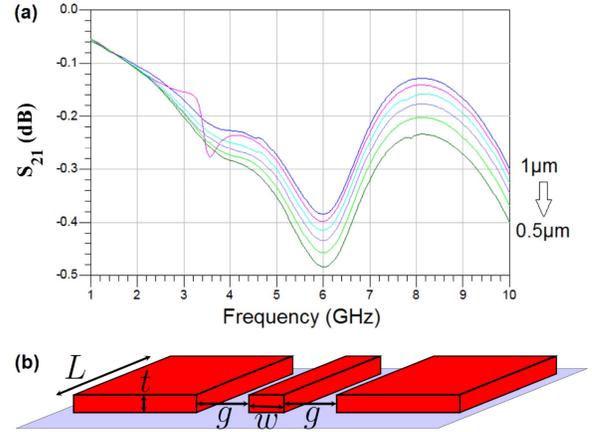

Fig. 2. (a) Simulations of a CPW with length $L = 2.5$ cm, central track width $w = 0.65$ mm, gap $g = 0.25$ mm, and varying thickness $t$. (b) Schematics of a CPW with the definition of $t$, $w$, $L$, and $g$.

We have performed high-frequency (HF) measurements on a system made up of two such probes and a set of three parallel copper strips. The probes were facing each other, their tongs either in direct or indirect contact via the underlying copper strips, as shown in Fig. 1(b) (only one probe is present for the sake of clarity). The resulting quadrupole's behavior was investigated with a network analyzer, yielding the curves of Fig. 1(c). $S_{11}$ and $S_{21}$ are the respective moduli of the two relevant $S$-matrix parameters [6]; they characterize the device by measuring the ratio of an output-to-input voltage as a function of input frequency. $S_{21}$ is measured after transmission through the whole device, while $S_{11}$, as a one-port measurement, quantifies the signal reflection at device's termination. Despite the fact that the copper strips are not optimized for a 50 Ω matching and that not one but two probes are seen, the results are satisfactory. Indeed, $S_{21}$ is above −3 dB up to 8 GHz, which is sufficient considering our pulsewidth. Other measurements (not shown here) in the time domain were carried out, where the pulses of 100 to 500 ps were sent either directly to an oscilloscope or through the aforementioned quadrupole. All of them were in agreement with the frequency domain experiments: the pulse shape was hardly affected, and its amplitude barely reduced (<4% reduction for the shorter pulse). We have also measured the transmission behavior of a prototype CPW [Fig. 1(c) (red line)] in a geometry suited for synchrotron experiments.

### B. Simulations of the Waveguide

Before resorting to full electromagnetic simulations, we used the free software TXLine to have good starting values for the geometry of the waveguide. In terms of length, the limitation mostly comes from the experiments in which to use the CPW (see Sections II-A and IV-B). Following [7], we found again a good line matching to 50 Ω characteristic impedance for the values of $w/(w + 2g)$ close to 1/2, where $w$ is the width of the central track and $g$ is the gap spacing between the central and the ground tracks [see Fig. 2(b)]. The value of $g$ was set to 0.25 mm by choosing a value slightly higher than the ratio of the maximum pulse amplitude of our generator to the dielectric strength of air. Then, we performed the simulations of the device for different values of the thickness $t$. The system geometry is shown in Fig. 2(b), the results in terms of transmission properties are shown in Fig. 2(a). Given the duration of our pulses, such a flat behavior up to 10 GHz is more than satisfactory. The dip at 6 GHz can be shown to be an electric length effect: a standing wave within the device. Furthermore, the dependence over $t$ indicates that thicker waveguides are better suited. We explain the difference between the simulation and Fig. 1(c) with faults in the fabrication process (small copper pieces having fallen into the CPW gap), and expect an improvement of 3–5 dB for $S_{21}$ in the gigahertz range once this is solved, paving the way for our nanosecond pulses.

## C. Experimental Realization and Tuning

As shown above, better performance is expected from thicker waveguides. We choose to grow the device out of copper via high-rate triode sputtering [8] not only for this reason but also to avoid an additional preparatory step. Indeed, since our substrates are commercial, 0.5 mm-thick alumina wafers, electrodeposition of copper would require a prior deposition of another metal and a proper contacting on the three narrow tracks. The copper pattern is obtained using laser lithography and the lift-off technique.

In order to obtain a suitable termination impedance, we resort to the AC magnetron sputtering of a $Ni_{80}Cr_{20}$ strip prior to the deposition of copper. Nickel–chromium alloys are used in commercial HF calibration kits because of their high resistivity ($\sim 2 \times 10^{-6}$ $\Omega$m) and resilience. This strip is transverse to the waveguide and is continuous from one ground plane to the other through the central track. It reaches one millimeter into each ground plane so as to ensure optimal electric contact. Since the material is two orders of magnitude more resistive than copper, this extra length is irrelevant to the termination impedance value as it is short-circuited by the surrounding copper. The dimensions are chosen based on the measured resistivity of the material in order to obtain a DC resistance slightly lower than the target 50 $\Omega$. Corrections are undertaken using FIB to increase the resistance value.

Once this terminated waveguide is fabricated, two other similar waveguides are produced in order to perform complete HF measurements. The aim is to fully characterize the S-matrix of the waveguide as a transmission line. The first one is terminated not by a $Ni_{80}Cr_{20}$ strip but by a wide copper short-circuit. The other one is exactly the same as the terminated CPW, only without the $Ni_{80}Cr_{20}$ impedance; it serves as an open-circuit version of the waveguide. Therefore, we can perform full short-open-load-through measurements, since this last version can also be used to investigate the transmission properties of the CPW.

## III. SAMPLE PREPARATION

### A. Cylindrical Nanowires With Diameter Modulations

Experiments carried out on nanostrips and nanowires usually rely on a setup where the magnetic field pulse is generated by a second metallic wire transverse to the magnetic system as an Oersted field [9], [10]. This scheme is practical, because both the investigate systems and the excitating device can be fabricated by lithography, one on top of the other. However, it is not suitable for systems fabricated within templates, such as nanoporous alumina membranes [11], [12]. Using this method, we produce electrodeposited Permalloy ($Ni_{80}Fe_{20}$) nanowires, which are then freed from their template upon dissolving the membrane, and must be thereafter dispersed on a substrate. The stochastic character of this operation makes the patterning of an antenna on top of the nanowires impractical, not to mention the typical length of the wires, from several hundred nanometers to a few dozen micrometers. Therefore, we choose either to make the dispersion of wires on top of the waveguide or to micromanipulate some of these and lay them at a suitable location.

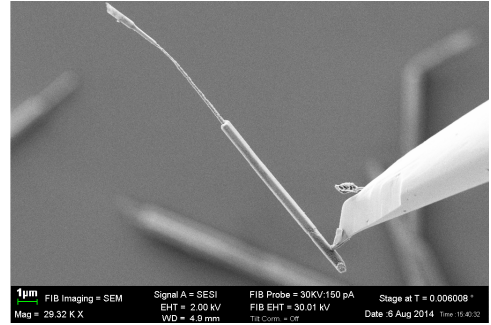

Fig. 3. FIB-assisted nanowire micromanipulation. The manipulating tip on the right barely contacts the wire, which has been lifted up off its substrate. Here, one of this wire's thick parts was broken during dispersion.

In terms of shape, we produce pores consisting of two thick parts ($\sim$150–200 nm in diameter) separated by a thinner part in the middle ($\sim$60–100 nm in diameter). Therefore, we benefit from the energy per unit surface of the DWs [1], and thus tailor barriers to prevent DWs from escaping. As a result, the risks of expelling DWs from the nanowires by applying too large fields are reduced.

### B. Micromanipulation

Nanowires are manipulated using a field-effect scanning electron microscope (model Leo 1530 by Zeiss) equipped with a micromanipulating tip. Upon closing in of the latter to a nanowire, a gallium-ion beam is used for bonding. The attaching area is chosen as close as possible to the wire end in order not to affect the region of interest, that is to say the thinner central part, where DWs can be trapped. The nanowire is lifted up from its initial dispersion substrate, as shown in Fig. 3, and laid on top of the waveguide. The bonding is then cut using the FIB. We have performed tests (not shown here) laying one nanowire on top of a copper surface much rougher than the sputtered waveguides: AFM measurements indicated a 200 nm root-mean-square roughness on a 10 × 10 $\mu m^2$ square. No alteration of the nanowire was visible.

## IV. SYNCHROTRON EXPERIMENTS

### A. XMCD-PEEM Imaging

XMCD is a synchrotron technique suited to investigating DWs because of its sensitivity to the magnetization orientation in 3d transition metals when the photon energy is chosen to match a transition from a core 2p electronic level to a 3d state at the Fermi level. The magnetization arises from these 3d states and their occupation; thus, with the selection rules constraint, the number of absorption events is affected by the net averaged magnetization at this location. This contribution to photon absorption is related to the orientation of magnetization with respect to the incoming circularly polarized X-ray beam. Upon switching from left circular polarization, or helicity, to right circular polarization, this contribution changes sign. Therefore, subtracting two images taken with opposite helicities yields a magnetic contrast image because all other contributions cancel out.

In our case, XMCD is coupled to PEEM for imaging. The transitions induced by the X-rays produce electrons,

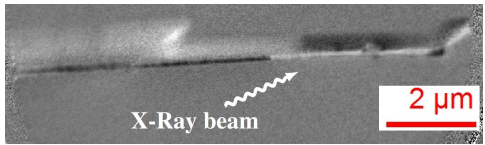

Fig. 4. XMCD image of a trisegmented nanowire and its shadow. There is an angle of ∼30° between the X-ray beam and the wire. A DW is pinned in the thin middle section of the wire, as shown by the abrupt contrast change.

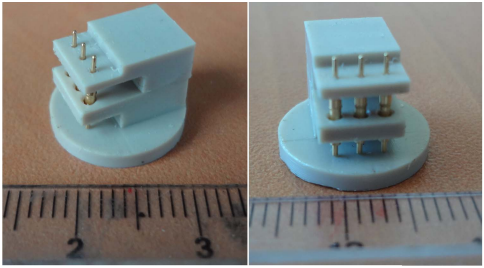

Fig. 5. Sample holder for synchrotron experiments with mounted spring-loaded contacts. The CPW is laid on the top flat surface and held in place by a metallic cap (not shown here) belonging to the microscope optics.

which are collected for imaging. Given the typical photon and electron mean free paths at the energies of interest, PEEM is a surface technique. If XMCD is not exploited, PEEM yields a plain electron microscopy image of the sample, whereas it gives a picture of the local magnetic configuration with the XMCD processing.

Identifying domains of opposite orientations in nanowires is possible with the X-ray beam along the wire, because the XMCD contrast is maximum when the magnetization and the photon wave vector are parallel. If a wire imaged with XMCD-PEEM displays two neighboring regions of opposite contrast, then it implies that the border between these is a DW. Then, imaging with the beam perpendicular to the nanowire gives an optimal insight on its structure at the cost of domain imaging.

Complementary to the information provided by the electrons emitted from the nanowire surface, volume information is available thanks to the photons traveling through the wire. The latter's shadow contains an XMCD contrast resulting from the inner magnetic configuration, which is shown in Fig. 4. However, this contrast is difficult to interpret. That is why we have developed a simulation code [3], [13] to compute the shadow contrast from a micromagnetic configuration.

### B. Development of a Suitable Sample Holder

The constraints for synchrotron experiments are harsh primarily because of the very limited space within the microscope on the Nanospectroscopy beamline at Elettra [14]. In addition, both the sample and the sample holder must be high vacuum-compatible. Moreover, since the pulses are coming in toward the sample from the bottom of the sample holder, which is shown in Fig. 5, it is necessary to use a different method for contacting the CPW. We solve this problem using three spring-loaded contacts. They serve as vertical current inlets: on top, they will be soldered to the three CPW tracks, and at the bottom, the spring-loaded contacts will push against horizontal copper tracks (not shown for clarity), the latter being connected to the pulse generator via a coaxial cable.

Establishing a more direct link between the CPW and the coaxial cable is impossible because of the presence of a metallic cap (not shown here) on top of the sample holder. Given the lengths involved, the effect on the signal is negligible. This cap is used to provide an additional voltage to the sample with respect to the 20 kV of the electron microscope, and also contributes to its optics. The sample holder itself is held in place in a cartridge specifically designed to accommodate the surroundings within the microscope. This cartridge sets the upper limit for the sample size (roughly $1 \times 1$ cm$^2$), and therefore the waveguide length that must be much shorter than the MFM experiments.

ACKNOWLEDGMENT

This work was supported by the EU Seventh Framework Programme (FP7/2007-2013) through the M3d Project under Grant 309589. The authors would like to thank N. Dempsey, A. Dias, C. Hoarau, P. David, B. Fernandez, and J.-F. Motte for their assistance. They would also like to thank S. Pizzini and J. Vogel for their helpful advice and for fruitful discussions.